\documentclass[12pt,a4paper]{article}
\usepackage[hmargin=2.5cm,vmargin=3cm,bindingoffset=0cm]{geometry}
\usepackage[T1]{fontenc}
\usepackage{lmodern,amsfonts,amsmath,amsthm,amssymb,mathrsfs,mathtools,hyperref}
\usepackage{chngcntr}
\usepackage[affil-it]{authblk}

\hypersetup{
    unicode=true,
    pdffitwindow=false,
    pdfstartview={FitH},
    pdftitle={Generalizations of Divisor Methods to Approval and Score Voting},
    pdfnewwindow=true,
    colorlinks=false,
    linkcolor=red,
    citecolor=blue,
    filecolor=blue,
    urlcolor=blue
}

\theoremstyle{definition}\newtheorem{dfn}{Definition}[section]
\theoremstyle{definition}\newtheorem{alg}{Algorithm}
\theoremstyle{definition}\newtheorem{ex}{Example}[section]
\theoremstyle{definition}\newtheorem{rem}{Remark}[section]
\theoremstyle{plain}
\theoremstyle{plain}
\theoremstyle{plain}
\theoremstyle{plain}\newtheorem{prop}{Proposition}[section]

\newenvironment{pf}{\noindent{\bfseries Proof }}{\hfill\qed\vspace*{.5em}}
\numberwithin{equation}{section}

\DeclarePairedDelimiter\ceil{\lceil}{\rceil}
\DeclarePairedDelimiter\floor{\lfloor}{\rfloor}

\newcommand{\comp}{\raisebox{.36pt}{\hspace{.2em}\scalebox{.8}{$\circ$}\hspace{.2em}}}
\newcommand{\st}{\,:\,}
\newcommand{\stl}{\,:\,}
\newcommand{\QQ}{\mathbb{Q}}
\newcommand{\RR}{\mathbb{R}}
\newcommand{\ZZ}{\mathbb{Z}}
\newcommand{\NN}{\mathbb{N}}
\newcommand{\normalize}[1]{\widetilde{#1}}

\title{Generalizations of Divisor Methods\\to Approval and Score Voting\vspace{2ex}}
\author{Martin Djukanovi\'{c}\thanks{The author is grateful to the University of Copenhagen for their hospitality. The author is also grateful to Warren D. Smith and Toby Pereira who provided useful guiding examples.}}
\affil{Mathematisch Instituut Leiden,\\ Institut de Math\'{e}matiques de Bordeaux}

\date{}

\begin{document}
\maketitle
\begin{abstract}
We introduce several electoral systems for multi-winner elections with approval ballots, generalizing the classical methods of Sainte-Lagu\"{e} and D'Hondt. Our approach is based on the works of Phragm\'{e}n~\cite{phragmen} and Thiele~\cite{thiele}. In the last section we discuss possible generalizations to score voting.
\end{abstract}

\vspace*{.5em}
\section{A brief overview of divisor methods}
Methods of largest quotients, also known as divisor methods,\footnote{We shall use these two names interchangeably throughout.} are methods of apportionment of seats in parliamentary democracies that use the so called \emph{party-list proportional representation}. We recall the description of such an electoral system:
\begin{itemize}
\item The number of seats in the parliament is fixed before the election, but it might vary from election to election.
\item Only entities known as {political parties} may participate in the election.
\item Each political party generates a list of its candidates. This list may be fully or partially public or completely private.
\item Each election ballot contains the list of political parties participating in the election and each voter must mark {exactly one} party on the ballot in order to cast a valid vote.
\item Based on the votes cast, a seat apportionment method proportionally allocates a number of seats to each party and the parties fill their seats sequentially according to their lists.
\end{itemize}

Let $I$ be a finite index set. Let $\{P_i\}_{i\in I}$ be the set of parties participating in the election and \mbox{let $n_i$} denote the number of votes that $P_i$ received. Further, let $n=\sum_{i\in I}n_i$ be the total number of votes (i.e. valid ballots) and let $s$ be the number of seats to apportion. The rational number $q_i={n_i}s/{n}$ is called $P_i$'s \emph{quota} and can be said to be this party's deserved portion of seats.\par
The following three are considered desirable properties that a proportional apportionment method should satisfy:
\begin{itemize}
\item[1)] (\emph{seat monotonicity}) If $s$ increases and the votes remain the same, no party should lose a seat in the new apportionment.
\item[2)] (\emph{vote monotonicity}) If $n_i$ increases whilst $n_j$ decreases, then $P_i$ should not lose a seat whilst $P_j$ does not lose a seat or gains a seat in the new apportionment.
\item[3)] (\emph{meeting quota}) Each party should receive either $\floor{q_i}$ or $\ceil{q_i}$ seats.
\end{itemize}
It has been known for a long time that divisor methods satisfy the two monotonicity properties but fail to meet quota. Balinski and Young \cite{by} showed that no method can satisfy both 2) and 3) and that only divisor methods satisfy both 1) and 2). They also constructed a method that satisfies both 1) and 3).\par
Monotonicity in votes is generally considered more important than meeting quota: an increase in support ought to lead to an increase, not decrease, in the number of seats even if it means an occasional unfairness (allocating to some parties strictly less than their lower quota or strictly more than their upper quota). This is the main reason why divisor methods are so widespread and why quota methods have been abandoned in places where they had been used previously. Moreover, empirically, quota violations in divisor methods seem to be rarer than monotonicity violations in quota methods. We now recall the two equivalent definitions of divisor methods.

\begin{dfn}[Largest Quotients Method of Apportionment]
For $i\in I$, let $s_i$ denote the number of seats that have been allocated to $P_i$ thus far. Initially, set $s_i=0$. A \emph{largest quotients method} allocates the next available seat to the party whose quotient
$$
\frac{n_i}{f(s_i)}
$$
is maximal. Here $f\colon \RR_{\geq 0}\to\RR_{\geq 0}$ is a monotonically increasing concave function, fixed a priori.
\end{dfn}

The following, equivalent formulation is probably more intuitive. 
\begin{dfn}[Divisor Method of Apportionment]
For $i\in I$, let $q_i={n_i}s/{n}\in\QQ$ denote the quota of party $P_i$. A \emph{divisor method} rescales all quotas by a suitable $\alpha>0$ such that
$$
\sum_{i\in I}\mathbf{r}\left( \alpha {q_i} \right) = s
$$
for some fixed rounding function $\mathbf{r}\colon \RR_{\geq 0}\to \ZZ_{\geq 0}$ and allocates $\mathbf{r}( \alpha {q_i} )$ seats to party~$P_i$ for each $i\in I$.
\end{dfn}
\begin{rem}
As long as all of the $n_i$ are pairwise distinct, a suitable $\alpha$ can be found and the procedure is well defined.
\end{rem}
The five classical versions of the method use the following functions:
{\renewcommand{\arraystretch}{1.6}
\begin{center}
\begin{tabular}{c|ccccc}
$\#$ & 1 & 2 & 3 & 4 & 5\\
\hline
${f(x)}$ & $x$ & $x+1$ & $x+\frac{1}{2}$ & $\sqrt{x(x+1)}$ & \raisebox{-.5em}{\scalebox{1.4}{$\frac{2x(x+1)}{2x+1}$}}
\end{tabular}
\end{center}}
\par\noindent
The corresponding rounding functions in the divisor formulation can be written as
$$
\mathbf{r}(x)=
\begin{cases}
      \floor{x}&\text{if } x < g(x), \\
      \ceil{x}&\text{otherwise,}
\end{cases}
$$
where $g(x)$ is as follows:
{\renewcommand{\arraystretch}{1.6}
\begin{center}
\centering
\begin{tabular}{c|ccccc}
$\#$ & 1 & 2 & 3 & 4 & 5\\
\hline
${g(x)}$ & $-\infty$ & $\infty$ & \raisebox{-.5em}{\scalebox{1.4}{$\frac{\floor{x}+\ceil{x}}{2}$}} & $\sqrt{\floor{x}\ceil{x}}$ & \scalebox{1.4}{$\frac{2}{\frac{1}{\floor{x}}+\frac{1}{\ceil{x}}}$}
\end{tabular}
\end{center}}\par\noindent
In other words, after rescaling by an appropriate $\alpha$, the rounding function rounds~$x$ to either $\ceil{x}$ or $\floor{x}$; the first method rounds up, the second rounds down, whilst the remaining three round the number down if and only if it is smaller than the arithmetic mean, the geometric mean and the harmonic mean of the two nearest integers, respectively. Ossipoff \cite{ossipoff} has suggested \mbox{$g(x)=\exp(\ceil{x}\log\ceil{x}-\floor{x}\log\floor{x}-1)$,} albeit in a different context.\par  
The five classical methods are respectively known as the methods of Adams, Jefferson (a.k.a. D'Hondt), Webster (a.k.a. Sainte-Lagu\"{e}), Hill (a.k.a. Equal Proportions) and Dean.\footnote{These divisor methods were first used for apportioning seats in the US House of Representatives based on state populations and they are named after their inventors. They were later rediscovered in Europe, as methods of largest quotients, where they are known under different names.} The most commonly used variants are D'Hondt and Sainte-Lagu\"{e}. They are usually formulated as follows: divide all of the $n_i$ by a sequence of increasing numbers and allocate the $s$ seats to the parties that correspond to the $s$ largest quotients. The D'Hondt sequence is $1,2,3,4,\dots$, whereas the Sainte-Lagu\"{e} sequence is $1,3,5,7,\dots$. These can be modified easily to tweak the properties of the method; e.g. the sequence $1.4,3,5,7,\dots$ is used in some countries with the intent of making it more difficult for parties to win their first seat.\par
Note that the Adams method has an obvious bias towards smaller parties, whilst the Jefferson/D'Hondt method has an obvious bias towards larger parties. The D'Hondt method is nevertheless used throughout Europe (and elsewhere) for various reasons.\footnote{It encourages formation of large parties and coalitions, supposedly leading to more stability. It also meets the lower quota, awarding at least $\floor{q_i}$ seats to $P_i$.} The Sainte-Lagu\"{e} variant is arguably the fairest among these five because it sequentially minimizes the variance of the number of representatives per voter, making representations as even as possible. On the other hand, D'Hondt only minimizes the amount of representation of the voter with the most representation. Empirically, the Sainte-Lagu\"{e} method seems to violate quota less often than the other four methods.
\par
\section{Approval and score voting}
Consider a single-winner election with two candidates, $A$ and $B$. The best method of choosing the winner in this election is the obvious one -- the winner is the candidate with the majority of the votes (cf. May's theorem \cite{may}). This simple election can be generalized to elections with more than two candidates and those can be further generalized to elections with more than one winner. We briefly recall the different approaches to the first step in the generalization.
\begin{itemize}
\item The simplest generalization to the case of more than two candidates imposes the condition that the voter must choose exactly one candidate to support. The candidate with the most votes is the winner; this is known as \emph{plurality rule}. This method does not permit the voter to express that she does not consider the candidates she did not vote for to be equally bad or that she considers some of them to be at least as good as the one she did vote for, leading to well known problems in practice, such as dishonestly choosing ``the lesser of two evils'' so as not to split the vote against the greater evil.
\item One approach, which may be called the Condorcet approach, interprets a vote for~$A$ instead of~$B$ as a statement of ``preference for $A$ over $B$'', denoted $A\succ B$. If one increases the number of candidates, this is then generalized to a preference chain $A \succ B \succ C \succ\dots$. The winner of such an election is declared to be the candidate who would pairwise beat any other candidate (in an election comparing preferences for one over the other), if such a candidate exists. Such a candidate is called the Condorcet winner and a method that always elects the Condorcet winner, if one exists, is called a \emph{Condorcet method}. Many such methods have been developed and they differ in how they treat the case when there is no Condorcet winner. While they satisfy many interesting properties, they are not free of paradoxes (cf. Arrow's theorem \cite{arrow}) and they are usually computationally demanding. We consider Condorcet methods too deficient to be used for electing state officials and we do not discuss them further.
\item Another generalization interprets a ballot in the $A$ vs $B$ election as splitting the candidates into two subsets, namely those that the voter approves and those that the voter does not approve. This has an obvious generalization known as \emph{approval voting} -- the voter marks the candidate(s) she approves and the most approved candidate wins.
\item One can also interpret a ballot in the $A$ vs $B$ election as giving the worst candidate a score of zero and giving the best candidate a score of one.
This is then generalized to giving every candidate some score $t\in[0,1]$ and is known as \emph{score voting} (or \emph{range voting}). The candidate with the highest total score wins the election. Not every possible score may be given in practice, however. Instead, for simplicity, the only scores that may be awarded are $$t\in\left\{0,\frac{1}{N},\frac{2}{N},\dots,\frac{N-1}{N},1\right\}\subset [0,1]$$
for some fixed $N\in\NN$. The method can be modified slightly to allow votes of no opinion, but this is not something we shall be dealing with in this paper.
\end{itemize}
\begin{rem}
In Score, there is an incentive for voters to maximize the impact of their vote by strategically awarding only scores of either zero or one. Since not all voters would use this to their advantage, it is sometimes argued that zero and one should be the only scores allowed, i.e. that Approval should always be used instead of Score. We contend that voters should always have the choice of awarding ``suboptimal'' scores.
\end{rem}\par
There are various other, exotic methods, but we do not discuss them here. Advantages and disadvantages of each are discussed at length elsewhere. We hold that score voting has numerous advantages over other methods, in terms of application to elections of state officials. A detailed list of these can be found at~\cite{website}. Whether or not the reader agrees with this view, it is the reason behind our endeavor to generalize Score to multi-winner elections. Our goal in this paper is to introduce particular natural generalizations that combine Approval/Score with the method of largest quotients. We wish for them to be scrutinized, improved upon, and compared with other methods.\par

\section{A generalization of Sainte-Lagu\"{e} to Approval}
We abandon entirely the party-list approach and consider an election with approval ballots in which the voter is free to mark any candidate she approves, regardless of party affiliation. From now on, $m$ shall denote the number of candidates, $n$ shall denote the number of voters, and $s$ shall denote the number of seats to apportion. The number of seats is essentially irrelevant because the method orders the candidates independently of this number\footnote{However, one can choose divisors that depend on the number of seats (see \cite{newappo}) or even the number of allocated seats.} and then fills the seats sequentially, according to the ordering. The ballots can be considered to be given in the form of an $m\times n$ binary matrix $(x_{ij})$ where
\[
    x_{ij}=
\begin{cases}
    1  & \text{if the $i$-th candidate was approved by the $j$-th voter},\\
    0 & \text{otherwise.}
\end{cases}
\]
We refer to the rows of this matrix as \emph{candidates}, denoting them by $x_1,x_2,\dots, x_m$, and we refer to the columns as \emph{voters}. Identical candidates are referred to as \emph{clones}.
\par
After permuting the columns if necessary, the usual ``choose one party-list'' election corresponds to a matrix of the following form:
\begin{equation*}
\left[ \begin{array}{ccccccccccccc}
1 & 1 & \cdots & 1 & & &  &  &  &  &  & & \\
\vdots & \vdots & \ddots & 1 &  &  &  &  &  &  &  & &  \\
1 & 1 & \cdots & 1 &  &  &  &  &  &  &  & &  \\
 &  &  &  & 1 & 1 & \cdots & 1 &  &  &  & &  \\
 &  &  &  & \vdots & \vdots & \ddots & 1 &   &   &  & &  \\
 &  &  &  & 1 & 1 & \cdots & 1 &   &   &   & &   \\
 &  &  &  &  &  &  &  & \ddots   &  &  & &   \\
 &  &  &  &  &  &  &  &  &  1 & 1 & \cdots & 1 \\
 &  &  &  &  &  &  &  &  &  \vdots & \vdots & \ddots & 1\\
 &  &  &  &  &  &  &  &  &  1 & 1 & \cdots & 1
\end{array} \right]
\end{equation*}
where the omitted values are all zeros. In other words, in the usual scenario, any two candidates are forced to be either clones or orthogonal. Recall that Sainte-Lagu\"{e} awards the next available seat to the candidate $x_i$ whose quotient
$$
\frac{n_i}{2s_i+1}
$$
is maximal. The integer $n_i$ is the number of votes for $x_i$ (and all its party clones) and may be denoted by $$|x_i|=\sum_{j=1}^n x_{ij}.$$
The integer $s_i$ is the number of clones of $x_i$ that have already been elected and it can be interpreted as the number of seats that are already representing the supporters of~$x_i$. Removing the choose-one-party-list restriction and allowing the ballot matrix to be any binary matrix of size $m\times n$, we can generalize $n_i$ and $s_i$ in a natural way. If a candidate~$x_i$ is elected and $|x_i|=n_i$, we consider that the seat of $x_i$ is equally split among the $x_i$-voters; each one is considered to be represented by ${1}/{n_i}$ of the seat.
This leads us to the definition of Algorithm~\ref{alg:alg1}.\par
We fix some notations first. We shall denote by $\langle\_,\_\rangle\colon \RR^n\times\RR^n\to\RR$ the usual scalar product, whilst 
\begin{align*}
|\_|\colon\RR^n\to\RR,\\
||\_||\colon\RR^n\to\RR
\end{align*}
shall denote the $L^1$- and $L^2$-norms, respectively. For a finite set $S$ of non-zero points in $\{0,1\}^n$, we shall denote by $\omega_S$ the sum of their normalizations with respect to the $L^1$-norm, that is
$$
\omega_S := \sum_{x\in S} \frac{x}{|x|}.
$$
If $S=\emptyset$ then $\omega_S = (0,0,\dots,0)$ by convention. We shall use the words \emph{set} and \emph{list} interchangeably when referring to the elected candidates as it is not always relevant that the candidates are elected in a specific order.
\begin{alg}[{``Phragm\'en-Sainte-Lagu\"{e}''}]
\label{alg:alg1}
Let $\mathcal{L}$ be the empty list. As long as there are candidates not on the list, append to it the candidate $x\notin\mathcal{L}$ with the maximal quotient
$$
\frac{|x|}{2\langle \omega_\mathcal{L} , x \rangle+1}.
$$
The list $\mathcal{L}$ gives the ordering of the candidates. Ties are broken by index (which is assumed to be pseudo-random in practice).
\end{alg}\par
Note that since $\omega_\emptyset={0}$, the first candidate on the list is always the most approved one. This algorithm is precisely the greedy algorithm that sequentially minimizes the variance of the number of representatives per voter for each seat it allocates (Proposition~\ref{thm:SL1}). When applied to the party-list scenario, this is just the usual Sainte-Lagu\"{e} method.\par
\begin{prop}
\label{thm:SL1}
Algorithm~\ref{alg:alg1} minimizes the variance of voter representation for each seat it allocates (assuming candidates with zero approval are ignored).
\end{prop}
\begin{pf}
Assume that there are no candidates with zero approval. 
 Voter representations are given at each step by the vector $\omega_\mathcal{L}$ and its variance is, by definition, the square of its $L^2$-distance from the ideal representations (that are given by the vector with the arithmetic mean of $\omega_\mathcal{L}$ in each coordinate), divided by $n$, i.e. if $k$ candidates have been elected then
$$
\text{Var}(\omega_\mathcal{L})=\frac{1}{n}\left\lVert\omega_\mathcal{L}-k\left(\frac{1}{n},\frac{1}{n},\dots,\frac{1}{n}\right)\right\rVert^2.
$$
Suppose, without loss of generality, that $\mathcal{L}=\{x_1,x_2,\dots,x_k\}$. Then it is readily seen that
\begin{equation*}
n\text{Var}(\omega_\mathcal{L})=\sum_{i=1}^k \frac{||x_i||^2}{|x_i|^2}-\frac{k^2}{n}+2\sum_{\substack{i,j=1 \\ i\neq j}}^k\frac{\langle x_i,x_j\rangle}{|x_i| |x_j|}.
\end{equation*}
It follows that $\text{Var}(\omega_{\mathcal{L}\cup\{x\}})$ is minimal precisely when $x$ is chosen so that
$$
\frac{|x|}{2\langle \omega_\mathcal{L}, x \rangle + \frac{||x||^2}{|x|}}
$$
is maximal. Since $x\in\{0,1\}^n$, we have $||x||^2=|x|$ and the claim follows.
\end{pf}
\par\par
This method has some obvious issues. One could say that it considers voter equality more important than candidate quality (i.e. popularity). Since we allow non-identical candidates to have common supporters, it is difficult to say to whom a particular candidate ``belongs'' -- a candidate can be shared by different groups. This can lead to a situation where the method considers a ``moderate'' candidate, shared between two groups, worse than an ``extreme'' candidate, supported by just one of the groups, because electing the moderate might over-represent one of the groups. This can be thought of as a kind of Pareto inefficiency: a candidate $x$ is elected even though there is a more popular candidate~$y$ that is approved by all (or most) of the $x$-voters. We elaborate on this in the examples that follow.
\begin{ex}
\label{ex:example1}
Let $k\geq 2$ be an integer. Consider the following distribution of approvals in an election:
\begin{align*}
k\text{ voters: } & A\\
1\text{ voter: } & A, B\\
k-1\text{ voters: } & B, C\\
\end{align*}
\noindent
Algorithm~\ref{alg:alg1} gives the ordering $(A,C,B)$ because allocating the second seat to~$B$ would give the $\{A,B\}$-voter too much representation; $C$ is orthogonal to~$A$, so $\{A,C\}$ represents the voters more evenly. This is in spite of the fact that $\{A,B\}$ is a \emph{Pareto improvement} over $\{A,C\}$; every $C$-voter is also a $B$-voter but not the other way around.
\end{ex}
Would it be a good idea to give the second seat to $B$? That depends on whether or not one considers voter approval (and voter representation to an extent) as a sort of welfare in the sense that increasing it for some does not harm others. On one hand, one can argue that if a group is over-represented, the parliament is biased. On the other hand, it counters intuition that some unpopular candidates should be overlooked only because some of their supporters also support some popular candidates. We hold it reasonable that small groups should not be discouraged from supporting some popular candidates over others in order to, as it were, swing the first allocated seats towards their side.
\par
\begin{rem}
A method cannot be considered fair if it places too much importance on popularity. In the extreme case, we could have a plurality-type method that simply awards all of the seats to the clones of the most popular candidate, provided that there are sufficiently many of them running. The same is true at the other end of the spectrum. Consider an example such as:
\begin{align*}
1\text{ : } & U, A_1\\
1\text{ : } & U, A_2\\
&\;\;\vdots\\
1\text{ : } & U, A_{k-1}\\
1\text{ : } & U, A_k
\end{align*}
If we are to apportion exactly $k$ seats, then the most proportional outcome would be to elect the $A_i$. However, this completely ignores the fact that $U$ is universally approved. It is therefore not a good idea to place too much importance on proportionality either.
\end{rem}
\par
The following is an example in a spirit similar to Example~\ref{ex:example1}.
\begin{ex}
\label{ex:phragmen}
Let $k\geq 1$ be an integer. Consider the following distribution of approvals:
\begin{align*}
20k\text{ : } & A, B, C\\
10k\text{ : } & X, Y, Z\\
2k\text{ : } & A, B, X\\
k\text{ : } & A, X, Y
\end{align*}
Algorithm~\ref{alg:alg1} gives $(A, X, C, B, Y, Z)$, regardless of $k$, even though $B$ is a Pareto improvement over $C$. One could alter the method so that it checks for Pareto improvements at each step. However, if we suppose that $k$ is large and we add one voter who approves only~$C$, we have $2k$ voters who prefer $B$ to $C$ and only one voter with the opposite preference. Then, even \mbox{though $B$} is no longer a Pareto improvement over~$C$ in the strict sense, one is tempted to elect $B$ over $C$ and argue that the opinion of the $2k$ voters matters more than the opinion of the single voter with the opposite preference. The question is, how many supporters of~$C$ need to be added before this decision is reversed?
\end{ex}

In the next section, we present a method that aims to deal with this issue.
\section{Improving the simple generalization}
The usual party-list divisor methods are a special case in which the number of voters who support candidate $x_i$ but not candidate $x_j$ is either $|x_i|$ or zero. Another way of describing the election procedure in that scenario is as follows. Consider all possible pairs $(x_i,x_j)$ of candidates that have not been awarded a seat thus far and define a relation $x_i \succ x_j$ for each step in the algorithm, i.e. for each seat to be filled, by the condition
$$
\frac{n_{ij}}{2s_{ij}+1} > \frac{n_{ji}}{2s_{ji}+1},
$$
where $n_{ij}$ is the number of voters who approve $x_i$ but not $x_j$ and $s_{ij}$ is the number of seats that represent these voters. The next available seat is given to a candidate $x$ such that $x\succ y$ for every $y\neq x$. Ties occur when $x=y$, when both quotients are zero, and they are broken by the party-list. Perhaps one could generalize to Approval from this point of view by defining the relation $x_i \succ x_j$ by the condition
$$
\frac{|\delta(x_i,x_j)|}{2\langle \omega_\mathcal{L}, \delta(x_i,x_j) \rangle+1} > \frac{|\delta(x_j,x_i)|}{2\langle \omega_\mathcal{L}, \delta(x_j,x_i) \rangle+1}
$$
with $\mathcal{L}$ being the set of candidates elected thus far and
\begin{equation}
\label{eq:delta}
\delta(a,b):=(\max\{0,a_k-b_k\})_k,\quad k=1,\dots,n.
\end{equation}
In other words, $\delta(x_i,x_j)\in\{0,1\}^n$ and its $k$-th coordinate is given by
\begin{equation*}
\delta(x_i,x_j)_k=
\begin{cases}
      1 & \text{if the }k\text{-th voter approves } x_i\text{ but not }x_j,\\
      0 & \text{otherwise.}
\end{cases}
\end{equation*}
However, this relation is not transitive in general and cycles can occur.
 One could consider applying a Condorcet method at this point, but we do not pursue that idea. Instead, we define the following method.
\begin{alg}[{``Pareto-improved Phragm\'en-Sainte-Lagu\"{e}''}]
\label{alg:alg2}
Let $\mathcal{L}$ be the empty list and let $\mathcal{C}$ denote the set of candidates. As long as there are candidates not on the list, do the following:
\begin{itemize}
\item[1)]Find the candidate $x\notin\mathcal{L}$ with the maximal quotient
$$
\frac{|x|}{2\langle \omega_\mathcal{L} , x \rangle+1}.
$$
\item[2)]If the set
$$S_x=\left\{z\in\mathcal{C}\backslash (\mathcal{L}\cup\{x\})\stl \frac{|\delta(z,x)|}{2\langle \omega_\mathcal{L} , \delta(z,x) \rangle+1}>\frac{|\delta(x,z)|}{2\langle \omega_\mathcal{L} , \delta(x,z) \rangle+1} \right\}$$
is empty, then append $x$ to $\mathcal{L}$, otherwise append $y\in S_x$ for which the difference
\begin{equation}
\label{eq:expr1}
\frac{|\delta(y,x)|}{2\langle \omega_\mathcal{L} , \delta(y,x) \rangle+1}-\frac{|\delta(x,y)|}{2\langle \omega_\mathcal{L} , \delta(x,y) \rangle+1}
\end{equation}
is maximal.
\end{itemize}
The list $\mathcal{L}$ gives the ordering of the candidates. A tie in step 1) is broken by comparing the quotients defined in step 2) (in favour of $x$ rather than $x'$ if $\delta(x,x')$ gives a larger quotient than $\delta(x',x)$) and if the quotients in step 2) are equal, the tie is broken by index.
\end{alg}
\par
In particular, a candidate $x$ will always be considered worse than a candidate that Pareto-dominates $x$, if such a candidate exists. When applied to Example~\ref{ex:phragmen}, this method gives the ordering $$(A, X, B, C, Y, Z)$$ and it takes exactly $\floor{\frac{598}{443}k}+1$ additional voters who approve only $C$ to change this ordering back to $$(A, X, C, B, Y, Z).$$\par
However, this method is not free of issues. The aim of step 2) is to improve upon the candidate from step 1). Of course, it matters which candidate the improvement is based upon and it might happen that one group of voters ``hijacks'' a candidate from another group.
\begin{ex}
\label{ex:counter_example}
Let the following matrix represent the distribution of approvals in an election.
\[
\begin{bmatrix}
a_1\\
a_2\\
a_3\\
a_4\\
b_1\\
b_2\\
b_3\\
b_4
\end{bmatrix}=
\begin{bmatrix}
1 & 1 & 1 & 1 & 1 & 1 & 1 & 0 & 1 & 1 & 0 & 1\\
1 & 1 & 1 & 1 & 0 & 1 & 1 & 1 & 0 & 1 & 0 & 0\\
1 & 1 & 1 & 1 & 1 & 1 & 0 & 1 & 0 & 1 & 0 & 0\\
1 & 1 & 1 & 1 & 0 & 1 & 0 & 1 & 0 & 1 & 0 & 0\\
0 & 0 & 0 & 0 & 1 & 1 & 0 & 1 & 1 & 1 & 1 & 0\\
0 & 0 & 0 & 0 & 1 & 1 & 1 & 1 & 0 & 1 & 1 & 0\\
0 & 0 & 0 & 0 & 0 & 1 & 0 & 0 & 0 & 1 & 1 & 1\\
0 & 0 & 0 & 0 & 0 & 0 & 0 & 0 & 0 & 0 & 0 & 1
\end{bmatrix}
\]
The first four voters approve only the $a_i$ candidates and, from their perspective, the two groups of candidates can be thought of as parties. The method awards the first seat to~$a_1$, it being the most popular candidate, with ten approval votes. We accordingly set
$$
\omega = \frac{a_1}{|a_1|}= \left(\frac{1}{10},\frac{1}{10},\frac{1}{10},\frac{1}{10},\frac{1}{10},\frac{1}{10},\frac{1}{10},0,\frac{1}{10},\frac{1}{10},0,\frac{1}{10}\right).
$$
The next seat goes to a candidate $x$ that maximizes
$$
\frac{|x|}{2\langle \omega,x \rangle+1}.
$$
There are three such candidates, namely $a_2$, $a_3$, and $b_1$. In the second step, $a_2$ and $a_3$ are tied and both beat $b_1$ so $a_2$ wins the second seat and we set
$$
\omega=\frac{a_1}{|a_1|}+\frac{a_2}{|a_2|}= \left(\frac{9}{40},\frac{9}{40},\frac{9}{40},\frac{9}{40},\frac{1}{10},\frac{9}{40},\frac{9}{40},\frac{1}{8},\frac{1}{10},\frac{9}{40},0,\frac{1}{10}\right).
$$
Step 2) of the algorithm does not give any improvements over $a_2$. Continuing in this way, the first four seats are awarded to $a_1$, $a_2$, $b_1$, $a_3$, in that order, without any improvements in step 2). Before the fifth seat is allocated, we have
$$
\omega=\left(  \frac{7}{20},\frac{7}{20}, \frac{7}{20}, \frac{7}{20},\frac{47}{120}, \frac{31}{60}, \frac{9}{40}, \frac{5}{12}, \frac{4}{15}, \frac{31}{60}, \frac{1}{6}, \frac{1}{10} \right)
$$
and
\begin{equation}
\label{eq:ineq1}
\frac{|b_3|}{2\langle \omega,b_3 \rangle+1}=\frac{10}{9}>\frac{45}{41}=\frac{|b_2|}{2\langle \omega,b_2 \rangle+1}>\frac{|a_4|}{2\langle \omega,a_4 \rangle+1}>\frac{|b_4|}{2\langle \omega,b_4 \rangle+1},
\end{equation}
so $b_3$ is the fairest choice for the fifth seat. Recall that the vectors
\begin{equation*}
\begin{split}
\delta(b_2,b_3)&=(0, 0, 0, 0, 1, 0, 1, 1, 0, 0, 0, 0)\\
\delta(b_3,b_2)&=(0, 0, 0, 0, 0, 0, 0, 0, 0, 0, 0, 1)
\end{split}
\end{equation*}
denote the voters with preferences $b_2 \succ b_3$ and $b_3 \succ b_2$, respectively. Since
$$
\frac{|\delta(b_2,b_3)|}{2\langle \omega,\delta(b_2,b_3) \rangle+1}=\frac{45}{46}>\frac{5}{6}=\frac{|\delta(b_3,b_2)|}{2\langle \omega,\delta(b_3,b_2) \rangle+1}
$$
and $b_2$ is the only available candidate for which such an inequality holds with respect to $b_3$, candidate $b_2$ is elected next. This is in spite of the fact that according to~\eqref{eq:ineq1}, electing $b_3$ is considered more proportional. We could say that candidate $b_2$ almost Pareto-dominates candidate~$b_3$; there is a single voter who prefers $b_3$ over $b_2$, but she is not sufficiently under-represented to justify electing $b_3$. Similarly, before the sixth seat is allocated we have
$$\omega = \left(\frac{7}{20},\frac{7}{20},\frac{7}{20},\frac{7}{20},\frac{67}{120},\frac{41}{60},\frac{47}{120},\frac{7}{12},\frac{4}{15},\frac{41}{60},\frac{1}{3},\frac{1}{10}\right)$$
and we check that
\begin{equation}
\label{eq:ineq2}
\frac{|a_4|}{2\langle \omega, a_4\rangle+1}= \frac{140}{87}>\frac{10}{7} =\frac{|b_3|}{2\langle \omega, b_3\rangle+1}>\frac{|b_4|}{2\langle \omega,b_4 \rangle+1}.
\end{equation}
However, we also have
$$
\frac{|\delta(b_3,a_4)|}{2\langle\omega,\delta(b_3,a_4)\rangle+1} = \frac{15}{14} > \frac{150}{149} = \frac{|\delta(a_4,b_3)|}{2\langle\omega,\delta(a_4,b_3)\rangle+1},
$$
and therefore candidate $b_3$ is elected next even though, according to \eqref{eq:ineq2}, electing~$a_4$ would be more proportional. However, if we suppose that only the first four voters voted for $a_4$, the situation changes. To illustrate the point, suppose that we added a candidate \mbox{$a_5=(1,1,1,1,0,0,0,0,0,0,0,0)$} to the election. Then $a_5$ would be considered the best at step 1) for the sixth seat. However, at step 2), candidate~$a_4$ would be considered an improvement over $a_5$ and $b_3$ would not, so $a_4$ would be elected instead.
\end{ex}
\begin{rem}
It is difficult to say to what extent this would be a problem in practice, however. It is certainly possible for a voter to approve certain winners, accumulate enough representation, and cast a vote of approval that harms a relatively unpopular candidate. In particular
$$
\frac{|x|+1}{2(\langle\omega_\mathcal{L},x\rangle+r)+1}<\frac{|x|}{2\langle\omega_\mathcal{L},x\rangle+1} \Leftrightarrow r>\frac{2\langle\omega_\mathcal{L},x\rangle+1}{2|x|}.
$$
Introducing step 2) with $\delta(x,y)$ helps the situation somewhat because $|x|\geq|\delta(x,y)|$. One idea worth considering might be the addition of placeholder \emph{phantom candidates}, such that each voter can vote for only one such candidate. These candidates would be present only so that they could be chosen in step~1) for improvement in step~2) and would be ineligible otherwise. The intended effect is similar to the effect of adding~$a_5$ in the example above. This could be accomplished by merging the ballot with the traditional choose-one-party ballot that contains the list of registered parties and the option \emph{None} available to the independents, who would not have a phantom candidate of their own. However, we are still left with the possibility of larger parties intentionally hijacking the phantom candidates of smaller parties.\end{rem}
\begin{rem}
The underlying philosophy of this method, and the sense in which it is a generalization of party-list divisor methods, can be stated as follows: \emph{determine which party is to receive the next seat and then give the seat to the best candidate of that party}. The word \emph{party} is understood as a group of like-minded voters, not candidates. The method itself determines the ``parties'' and the ``party-lists'', as it were. The variant with phantom candidates is guided towards  the actual parties. Other variants of the method can be obtained by replacing the corresponding divisors by the various other possibilities mentioned in the first section. Moreover, different divisors can be used in steps 1) and 2).
\end{rem}
\section{Thiele, Phragm\'{e}n and a new approach}
\emph{Phragm\'{e}n's method} \cite{phragmen}, properly interpreted, turns out to be the D'Hondt variant of the method described by Algorithm~\ref{alg:alg1}. It also uses the vector $\omega$, whose coordinates Phragm\'{e}n calls \emph{loads}, a name that is perhaps a better choice than ours. In this method, the candidate who wins the next available seat is the \mbox{candidate $x$} whose supporters would have the minimal average load after $x$ is elected. If~$\mathcal{L}$ is the list of candidates elected thus far, Phragm\'{e}n elects~$x$ that minimizes
$$
\frac{\langle \omega_{\mathcal{L}\cup\{x\}} ,x \rangle}{|x|},
$$
which is to say that $x$ maximizes
$$
\frac{|x|}{\langle \omega_{\mathcal{L}}+\frac{x}{|x|} ,x \rangle}=\frac{|x|}{\langle \omega_{\mathcal{L}},x \rangle+ \frac{1}{|x|}\langle x,x\rangle}=\frac{|x|}{\langle \omega_{\mathcal{L}},x \rangle+ 1},
$$
which is clearly a generalization of D'Hondt.\par
\emph{Thiele's method} \cite{thiele} works by assigning weights to the ballots. For each seat to be filled, each voter's ballot is first divided by $s_i+1$, where $s_i$ is the number of elected candidates that the $i$-th voter approved, and the seat is given to the candidate with the highest total score after summing the reweighted ballots. Instead of dividing by~$s_i+1$, it is also possible to define the method by divisors $2s_i+1$ etc. Note that Thiele's method also generalizes D'Hondt and Sainte-Lagu\"{e} to Approval (depending on which divisors are chosen). However, Thiele's method is deficient in one important way: it strongly encourages the voters not to approve the candidates who are most probably going to be elected even without their support. Withholding such unnecessary support would increase significantly the impact of their vote when electing less popular candidates. This issue is less pronounced with Phragm\'{e}n.\par
\begin{rem}[Nomenclature]
With the history in mind, we suggest that the classical method of Phragm\'{e}n, as defined here, be referred to as \emph{Phragm\'{e}n-D'Hondt} and that the method given by Algorithm~\ref{alg:alg1} be referred to as \emph{Phragm\'{e}n-Sainte-Lagu\"{e}} (or respectively as the D'Hondt and the Sainte-Lagu\"{e} variants of Phragm\'{e}n's method). The improved method based on ``difference quotients'', given by Algorithm~\ref{alg:alg2}, might be named after the differences or referred to as \emph{Pareto-improved Phragm\'en} (Sainte-Lagu\"{e} variant as given in the definition, with obvious modifications to obtain other variants). We suggest analogous names for the corresponding variants of Thiele's method.
\end{rem}\par
The key difference between Phragm\'en and Thiele lies in how they treat the notion of the amount of representation that a voter has and how that amount influences the election procedure. Under Thiele, if some voters approved an elected candidate, they are considered to be represented by that candidate. Under Phragm\'{e}n, they are considered to be represented by the candidate only to the extent that they contributed to getting said candidate elected. Note that Phragm\'{e}n can be interpreted as a method that reweights the votes of like-minded voters by the same amount and then compares the candidates' total scores. Both methods reweight the votes in a way that can be seen as unfair towards small groups. Thiele reweights the \emph{entire} ballot and does so without considering the popularity of elected candidates, whilst Phragm\'{e}n reweights a vote for a particular candidate on a voter's ballot based almost entirely on the previous choices of \emph{other} supporters of that candidate.
\par
\begin{ex}[Example~\ref{ex:counter_example} revisited]
If we put $a_1=(1, 1, 0, 0, 1, 1 , 1, 0 , 1 , 1, 0, 1)$, changing the votes of the third and the fourth voter, then the position of $a_4$ is improved under both Phragm\'{e}n and Thiele, in both the classical variants and the variants with differences (using $\langle\delta(x,y),r(x)\rangle$ for purposes of Pareto improvements in Thiele's method, where $r(x)$ denotes the candidate $x$ with coordinates reweighted with respect to the set of elected candidates using Sainte-Lagu\"e divisors).\par
\end{ex}

Note that Thiele always reweights the columns of the ballot matrix by a fixed amount, whereas Phragm\'en reweights the rows by a fixed amount (see also Example~\ref{ex:revis_1}). This suggests that a reasonable Approval method might be obtained by combining the two approaches. In what follows, we present one such method and its corresponding Pareto-improved variant.\par
Our goal is to reweight the entries in the ballot matrix individually instead of reweighting a row or a column by a fixed amount. We wish to do so in a way that only considers individual representations, not collective (like Phragm\'en), and a way that is not excessively harsh towards small groups (like Thiele is). 
\par
Consider again the case of the choose-one-party-list election, this time from the points of view of Phragm\'en and Thiele. If at some step in the election procedure a voter of~$P_i$ has representation ${s_i}/{n_i}$, that is simply to say that the party with $n_i$ votes has won $s_i$ seats thus far. For the purpose of reweighting, Thiele comes up with~$s_i$ as the number of summands in
$$
\frac{1}{n_i}+\dots+\frac{1}{n_i}=\frac{s_i}{n_i},
$$
counting how many times this voter has obtained some positive amount of representation. In the general case, Thiele reweights the entire ballot without taking into account the actual amounts. On the other hand, Phragm\'en comes up with~$s_i$ by summing the individual representations of the voters who approved the next available $P_i$-clone on the party list. These are precisely the $n_i$ voters with  ${s_i}/{n_i}$ representation each and we have
$$
\sum_{j=1}^{n_i}\frac{s_i}{n_i}=s_i.
$$
In the general case, Phragm\'en reweights all votes for a candidate equally, without taking into account the individual contributions to the total amount of representation that the supporters of the candidate have.
\par
To meet our goal, we suggest that $s_i$ be interpreted as the product of the current representation of the $P_i$-voter and the number of votes for the $P_i$-candidate:
\begin{equation}
\label{eq:eq1}
\frac{s_i}{n_i} n_i=s_i.
\end{equation}
Equivalently, this can be seen as the ratio of the amount of representation that the $P_i$-voter currently has and the amount that she would receive if one of the $P_i$-candidates was elected next, i.e.
\begin{equation}
\label{eq:eq2}
\frac{s_i}{n_i} / \frac{1}{n_i}=s_i.
\end{equation}
In the general case, dropping the $P_i$- prefix, this approach would treat like-minded groups corresponding to clones (resp. near-clones) much like Sainte-Lagu\"e, dividing precisely (resp. approximately) by $1,3,5,7,\dots$ as they got elected and, at the same time, it would not punish excessively any small groups that have partial agreement with larger groups. We introduce the necessary notation before writing down the formal definition.\par
For any $y\in[0,1]^n$ and $w\in\RR_{\geq 0}^n$ we denote by~$r(y,w)$
the point $y$ with coordinates reweighted with respect to $w$ in the sense above, based on~\eqref{eq:eq1}. That is to say that
\begin{equation}
\label{eq:newrew}
r(y,w):=\left(\dfrac{y_1}{2w_1{|y|}+1},\dfrac{y_2}{2w_2{|y|}+1},\dots,\dfrac{y_n}{2w_n{|y|}+1}\right).
\end{equation}
For later convenience, we also introduce
\begin{equation}
\label{eq:newrewdelta1}
r_\delta(y,z,w):=\left(\dfrac{\max\{y_1-z_1,0\}}{2w_1{|y|}+1},\dfrac{\max\{y_2-z_2,0\}}{2w_2{|y|}+1},\dots,\dfrac{\max\{y_n-z_n,0\}}{2w_n{|y|}+1}\right)
\end{equation}
for any $z\in[0,1]^n$. Note that for $y,z\in\{0,1\}^n$ we have $|r_\delta(y,z,w)|=\langle \delta(y,z) , r(y,w) \rangle$.
\begin{alg}
\label{alg:new}
Let $\mathcal{L}$ be the empty list and let $\mathcal{C}$ denote the set of candidates. As long as there are candidates not on the list, append to it the candidate $x\notin\mathcal{L}$ with the maximal norm
$
|r(x,\omega_\mathcal{L})|.
$
 The list $\mathcal{L}$ gives the ordering of the candidates. Ties are broken by index.
\end{alg}

\begin{alg}
\label{alg:new2}
Let $\mathcal{L}$ be the empty list and let $\mathcal{C}$ denote the set of candidates. As long as there are candidates not on the list, do the following:
\begin{itemize}
\item[1)]Find the candidate $x\notin\mathcal{L}$ for which the norm
$
|r(x,\omega_\mathcal{L})|
$
 is maximal.
\item[2)]If the set
$$S_x=\left\{z\in\mathcal{C}\backslash (\mathcal{L}\cup\{x\})\stl |r_\delta(z,x,\omega_\mathcal{L})| > |r_\delta(x,z,\omega_\mathcal{L})| \right\}$$
is empty, then append $x$ to $\mathcal{L}$, otherwise append $y\in S_x$ for which the difference
\begin{equation*}
|r_\delta(y,x,\omega_\mathcal{L})| - |r_\delta(x,y,\omega_\mathcal{L})|
\end{equation*}
is maximal.
\end{itemize}
The list $\mathcal{L}$ gives the ordering of the candidates. A tie in step 1) is broken by comparing the norms defined in step 2) (in favour of $x$ rather than $x'$ if $|r_\delta(x,x',\omega_\mathcal{L})|>|r_\delta(x',x,\omega_\mathcal{L})|$) and if the norms in step 2) are equal, the tie is broken by index.
\end{alg}
\par
Note that $r(x,\omega_{\mathcal{L}})$ reweights an individual vote for $x$ based only on the popularity of~$x$ and the individual representation of the voter, and that the method is free of artificial constructions such as phantoms etc. It is clear how other variants, such as D'Hondt, can be obtained by modifying~\eqref{eq:newrew} and~\eqref{eq:newrewdelta1}.\par
When applied to Example~\ref{ex:phragmen}, Algorithms~\ref{alg:new} and~\ref{alg:new2} give the same result as the two Phragm\'en algorithms, but now only~$\floor{\frac{598}{1883}k}+1$ additional $C$-voters are needed for Algorithm~\ref{alg:new2} to elect~$C$ over~$B$.
\begin{ex}[Example~\ref{ex:counter_example} revisited]
\label{ex:revis_1}
To see more clearly the difference between Thiele, Phragm\'en and Algorithm~\ref{alg:new} (all in Sainte-Lagu\"{e} variants), we take a look at the reweighted ballot matrix in each method when $\mathcal{L}=\{a_1,a_2,b_1,a_3\}$. In Thiele we have
\[
\begin{bmatrix}
\frac{1}{7} & \frac{1}{7} & \frac{1}{7} & \frac{1}{7} & \frac{1}{7} & \frac{1}{9} & \frac{1}{5} & 0 & \frac{1}{5} & \frac{1}{9} & 0 & \frac{1}{3}\\[.3em]
\frac{1}{7} & \frac{1}{7} & \frac{1}{7} & \frac{1}{7} & 0 & \frac{1}{9} & \frac{1}{5} & \frac{1}{7} & 0 & \frac{1}{9} & 0 & 0\\[.3em]
\frac{1}{7} & \frac{1}{7} & \frac{1}{7} & \frac{1}{7} & \frac{1}{7} & \frac{1}{9} & 0 & \frac{1}{7} & 0 & \frac{1}{9} & 0 & 0\\[.3em]
\frac{1}{7} & \frac{1}{7} & \frac{1}{7} & \frac{1}{7} & 0 & \frac{1}{9} & 0 & \frac{1}{7} & 0 & \frac{1}{9} & 0 & 0\\[.3em]
0 & 0 & 0 & 0 & \frac{1}{7} & \frac{1}{9} & 0 & \frac{1}{7} & \frac{1}{5} & \frac{1}{9} & \frac{1}{3} & 0\\[.3em]
0 & 0 & 0 & 0 & \frac{1}{7} & \frac{1}{9} & \frac{1}{5} & \frac{1}{7} & 0 & \frac{1}{9} & \frac{1}{3} & 0\\[.3em]
0 & 0 & 0 & 0 & 0 & \frac{1}{9} & 0 & 0 & 0 & \frac{1}{9} & \frac{1}{3} & \frac{1}{3}\\[.3em]
0 & 0 & 0 & 0 & 0 & 0 & 0 & 0 & 0 & 0 & 0 & \frac{1}{3}
\end{bmatrix},
\]
in Phragm\'en we have
\[
\begin{bmatrix}
\frac{6}{47} & \frac{6}{47} & \frac{6}{47} & \frac{6}{47} & \frac{6}{47} & \frac{6}{47} &\frac{6}{47} & 0   & \frac{6}{47} & \frac{6}{47} & 0   & \frac{6}{47}\\[.3em]
\frac{20}{143} & \frac{20}{143} & \frac{20}{143} & \frac{20}{143} & 0   & \frac{20}{143} & \frac{20}{143} & \frac{20}{143} & 0   & \frac{20}{143} & 0   & 0\\[.3em]
\frac{60}{449} & \frac{60}{449} & \frac{60}{449} & \frac{60}{449} & \frac{60}{449} & \frac{60}{449} & 0   &\frac{60}{449} & 0   & \frac{60}{449} & 0   & 0\\[.3em]
\frac{10}{67} & \frac{10}{67} & \frac{10}{67} & \frac{10}{67} & 0   & \frac{10}{67} & 0   & \frac{10}{67} & 0   & \frac{10}{67} & 0   & 0\\[.3em]
0   & 0   & 0   & 0   & \frac{20}{111} & \frac{20}{111} & 0   & \frac{20}{111} & \frac{20}{111} & \frac{20}{111} & \frac{20}{111} & 0\\[.3em]
0   & 0   & 0   & 0   & \frac{15}{82} & \frac{15}{82} & \frac{15}{82} & \frac{15}{82} & 0   & \frac{15}{82} & \frac{15}{82} & 0\\[.3em]
0   & 0   & 0   & 0   & 0   & \frac{5}{18} & 0   & 0   & 0   & \frac{5}{18} & \frac{5}{18} & \frac{5}{18}\\[.3em]
0   & 0   & 0   & 0   & 0   & 0   & 0   & 0   & 0   & 0   & 0   & \frac{5}{6}
\end{bmatrix},
\]
and in Algorithm~\ref{alg:new} we have
\[
\begin{bmatrix}
\frac{1}{8} & \frac{1}{8} & \frac{1}{8} & \frac{1}{8} & \frac{6}{53} & \frac{3}{34} & \frac{2}{11} & 0 & \frac{3}{19} & \frac{3}{34} & 0 & \frac{1}{3}\\[.3em]
\frac{5}{33} & \frac{5}{33} & \frac{5}{33} & \frac{5}{33} & 0 & \frac{15}{139} & \frac{5}{23} & \frac{3}{23} & 0 & \frac{15}{139} & 0 & 0\\[.3em]
\frac{5}{33} & \frac{5}{33} & \frac{5}{33} & \frac{5}{33} & \frac{15}{109} & \frac{15}{139} & 0 & \frac{3}{23} & 0 & \frac{15}{139} & 0 & 0\\[.3em]
\frac{10}{59} & \frac{10}{59} & \frac{10}{59} & \frac{10}{59} & 0 & \frac{30}{247} & 0 & \frac{6}{41} & 0 & \frac{30}{247} & 0 & 0\\[.3em]
0 & 0 & 0 & 0 & \frac{10}{57} & \frac{5}{36} & 0 & \frac{1}{6} & \frac{5}{21} & \frac{5}{36} & \frac{1}{3} & 0\\[.3em]
0 & 0 & 0 & 0 & \frac{10}{57} & \frac{5}{36} & \frac{10}{37} & \frac{1}{6} & 0 & \frac{5}{36} & \frac{1}{3} & 0\\[.3em]
0 & 0 & 0 & 0 & 0 & \frac{15}{77} & 0 & 0 & 0 & \frac{15}{77} & \frac{3}{7} & \frac{5}{9}\\[.3em]
0 & 0 & 0 & 0 & 0 & 0 & 0 & 0 & 0 & 0 & 0 & \frac{5}{6}
\end{bmatrix}.
\]\par
When applied to Example~\ref{ex:counter_example}, Algorithms~\ref{alg:new} and~\ref{alg:new2} respectively give
\begin{equation*}
\begin{split}
(a_1,b_1,a_2,a_3,b_3,a_4,b_2,b_4),\\
(a_1,a_3,b_1,a_2,b_2,b_3,a_4,b_4).
\end{split}
\end{equation*}
If we set $a_1=(1, 1, 0, 0, 1, 1 , 1, 0 , 1 , 1, 0, 1)$ then they respectively give
\begin{equation*}
\begin{split}
(a_1,a_2,b_1,a_3,b_3,a_4,b_2,b_4),\\
(a_1,a_2,b_1,a_3,b_2,a_4,b_3,b_4).
\end{split}
\end{equation*}
If $a_4=(1, 1, 1, 1, 0, 0 , 0, 0 , 0 , 0, 0, 0)$, Algorithm~\ref{alg:new2} gives $(a_1,a_3,b_1,a_2,b_2,b_3,a_4,b_4)$.
\end{ex}
\begin{rem}
Like Phragm\'en and Thiele, both of these algorithms are not monotonic; changing a vote from $0$ to $1$ can lower the position of the corresponding candidate, as can be seen by changing the first vote of the first voter in the following two ballot matrices:
\[
\begin{bmatrix}
0 & 0 & 0 & 1 & 1 & 0 & 1 & 1 & 0 & 0 & 1\\
0 & 0 & 0 & 0 & 1 & 1 & 0 & 0 & 0 & 1 & 0\\
0 & 1 & 0 & 1 & 1 & 0 & 0 & 1 & 0 & 1 & 0\\
0 & 1 & 0 & 1 & 0 & 1 & 0 & 1 & 0 & 0 & 0\\
1 & 0 & 0 & 0 & 1 & 1 & 0 & 1 & 1 & 0 & 0\\
1 & 1 & 0 & 1 & 0 & 1 & 1 & 0 & 0 & 1 & 1
\end{bmatrix}
\quad , \quad
\begin{bmatrix}
0 & 0 & 0 & 1 & 0 & 0 & 1 & 0 & 1 & 0 & 0\\
1 & 0 & 0 & 1 & 0 & 1 & 0 & 1 & 0 & 0 & 0\\
0 & 1 & 0 & 0 & 1 & 0 & 0 & 0 & 0 & 0 & 0\\
1 & 0 & 1 & 0 & 0 & 0 & 0 & 0 & 1 & 1 & 1\\
0 & 1 & 1 & 0 & 0 & 0 & 0 & 0 & 0 & 1 & 0\\
1 & 1 & 1 & 1 & 1 & 0 & 1 & 0 & 0 & 1 & 0
\end{bmatrix}.
\]
\end{rem}

\section{A geometric interpretation and generalizations to Score}
It is natural to ask whether one can generalize these methods to Score. One immediate generalization is obtained by converting the scores to approval votes. If the allowed scores in $[0,1]$ are integer multiples of $1/N$ for some $N\in\NN$, then we can treat every score voter as $N$ approval voters and convert the scores via
\begin{equation}
\label{eq:convert}
k/N \mapsto (\underbrace{1,1,\dots,1}_\textrm{$k$ ones},\underbrace{0,0,\dots,0}_\textrm{$N-k$ zeros}).
\end{equation}
However, one hopes to obtain a direct generalization by allowing votes between zero and one in Approval. We make a few observations.\par
There are infinitely many ways to continuously fill the gap between zero and one and it need not be done linearly. Unlike Approval, Score allows for the total score of a candidate~$x$ to arise in many different ways from the voters supporting $x$ (the voters who gave $x$ non-zero scores). The question is whether all these possibilities should be treated equally and whether the scores should simply be summed when comparing candidates. For example, one could argue that a hundred voters, each of whom contributed a score of $1/100$, should not be treated the same as one voter who contributed a score of 1. If these were the scores of two candidates, it could be that some of the hundred voters would be willing to change their scores to zero so that the voter with the opposite preference, who is much more passionate about her choice, can have her way. It is not clear how this should be treated. One might, for example, apply a transformation to the scores $t\in[0,1]$, such as $\vartheta(t)= t^2$ or $\vartheta(t)=\exp(1-\exp(\log(t)\log(a)^{-1}\log(1-\log(b))))$ with $a,b\in(0,1)$, chosen so that $\vartheta(a)= b$, etc.
\par
If a candidate $x$ is elected, one can argue that its contribution to voter representations should be given by ${x^2}/{|x|}$, with multiplication defined coordinate-wise. For example, if~$n$ voters give $x$ a score of $1/2$, each portion that is~$1/n$ of the filled seat can be said to be worth only $1/2$ of that amount to each voter, whence ${x^2}/{|x|}$. The same conclusion can be reached by considering the representations after the conversion given by~\eqref{eq:convert}.\par
Whether the representations are given by ${x}/{|x|}$ or ${x^2}/{|x|}$, a method that simply minimizes representation variance is not a good method because a candidate with a low score is considered good if the voters agree very much on how bad the candidate is. It is also not a good idea to simply minimize the distance from the ideal representation $u=(1/n,\dots ,1/n)$. For example, the distance \mbox{between $u$} and $(0,0,\dots,0)$ \mbox{approaches 0} as $n\rightarrow\infty$, whereas the distance \mbox{between $u$} and $(1,0,0,\dots,0)$ approaches 1. A consistent generalization should not only prefer lower variance in the number of seats per voter, it should also prefer higher candidate scores. Moreover, if we impose the condition that equal total scores should be considered equally good, all else being equal, this introduces additional difficulties.\par

\subsection{Generalizing Phragm\'en}
Consider again an election with approval votes. Phragm\'{e}n-D'Hondt and Phragm\'{e}n-Sainte-Lagu\"{e} can be interpreted as different takes on a particular optimization problem. Let us normalize the candidates with respect to the $L^1$-norm and set
$$
\normalize{x_i}:=\frac{x_i}{|x_i|},
$$
so that each $\normalize{x_i}$ is a point in the standard $(n-1)$-simplex
$$
\Delta^{n-1}=\left\{(t_1,t_2,\dots,t_n)\in\RR_{\geq 0}^{n}\stl \sum_{i=1}^n t_i =1 \right\}.
$$
Since $x_i\in\{0,1\}^n$, the normalized $\normalize{x_i}$ are actually midpoints of faces (of various dimensions) of the simplex $\Delta^{n-1}$. Clearly, the sum of any $k$ normalized candidates is a point in the rescaled simplex $k\Delta^{n-1}$. If we are given the set \mbox{$S=\{\normalize{x_1}, \normalize{x_2},\dots,\normalize{x_m}\}$}, we can interpret the Phragm\'{e}n method as the greedy algorithm that sequentially chooses points from $S$ in such a way that at the $k$-th step the distance between the sum~$\omega$ of the~$k$ chosen points and $\frac{k}{n}(1,1,\dots,1)$ is minimal. Since $k\Delta^{n-1}$ is orthogonal to the line $\ell=\{\alpha\cdot(1,1,\dots,1)\st\alpha\in\RR\}$, the same quality measure can be defined by the distance from $\ell$ or by the distance from some point $\alpha(1,1,\dots,1)\in\ell$. Note \mbox{that $\omega$} starts at the origin and moves through the cone defined by~$\Delta^{n-1}$ as the seats are filled. One hopes to generalize this method to Score by generalizing the normalization function and then defining a similar problem. If one were to use~$x^2/|x|$ as individual amounts of representation and minimize the distance between~$\omega$ and a point $\alpha(1,1,\dots,1)\in\ell$, it would not be clear \mbox{how $\alpha$} should be chosen and how it should change in different steps of the algorithm, if at all. A more distant reference point would discriminate more against lower scores.
\par
\begin{figure}[t]
\centering
\includegraphics[width=10em]{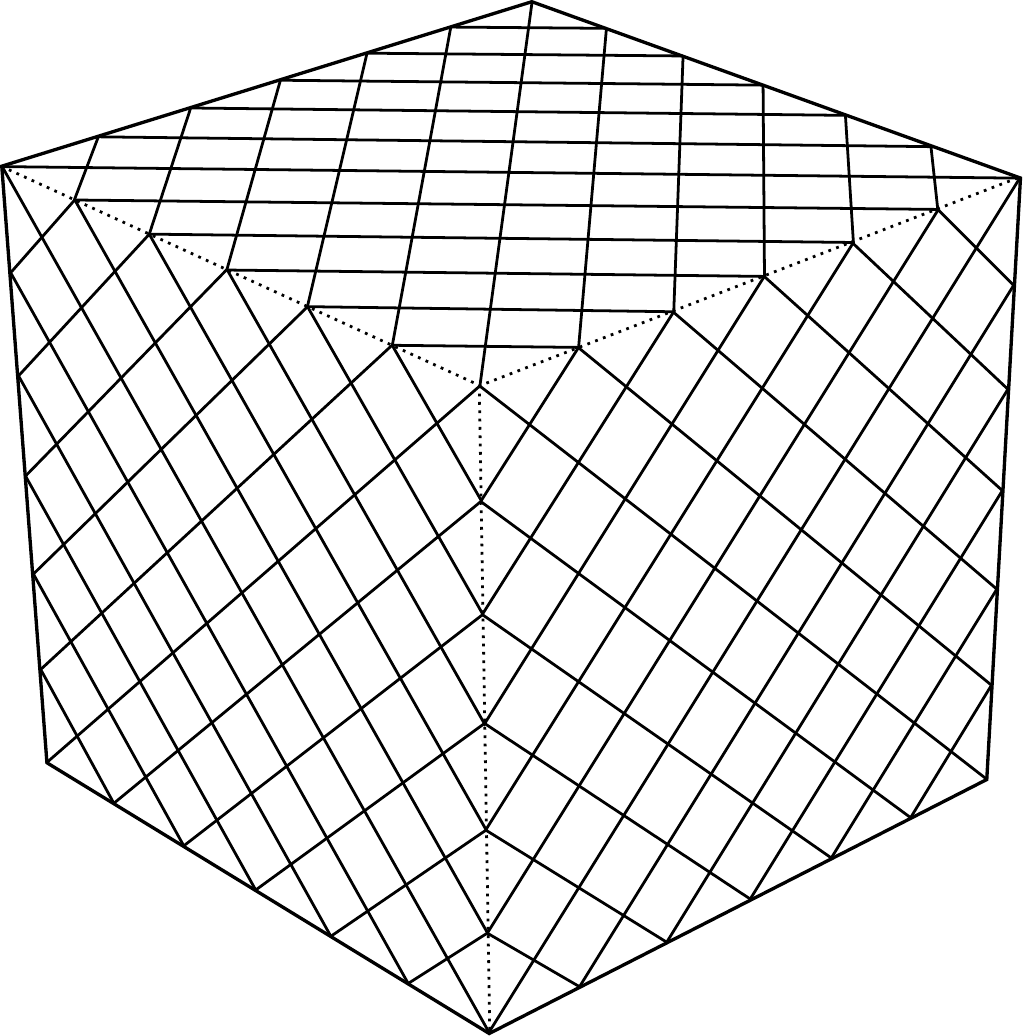}\qquad \raisebox{4em}{$\stackrel{\mu}{\longrightarrow}$}\qquad \raisebox{1em}{\includegraphics[width=10em]{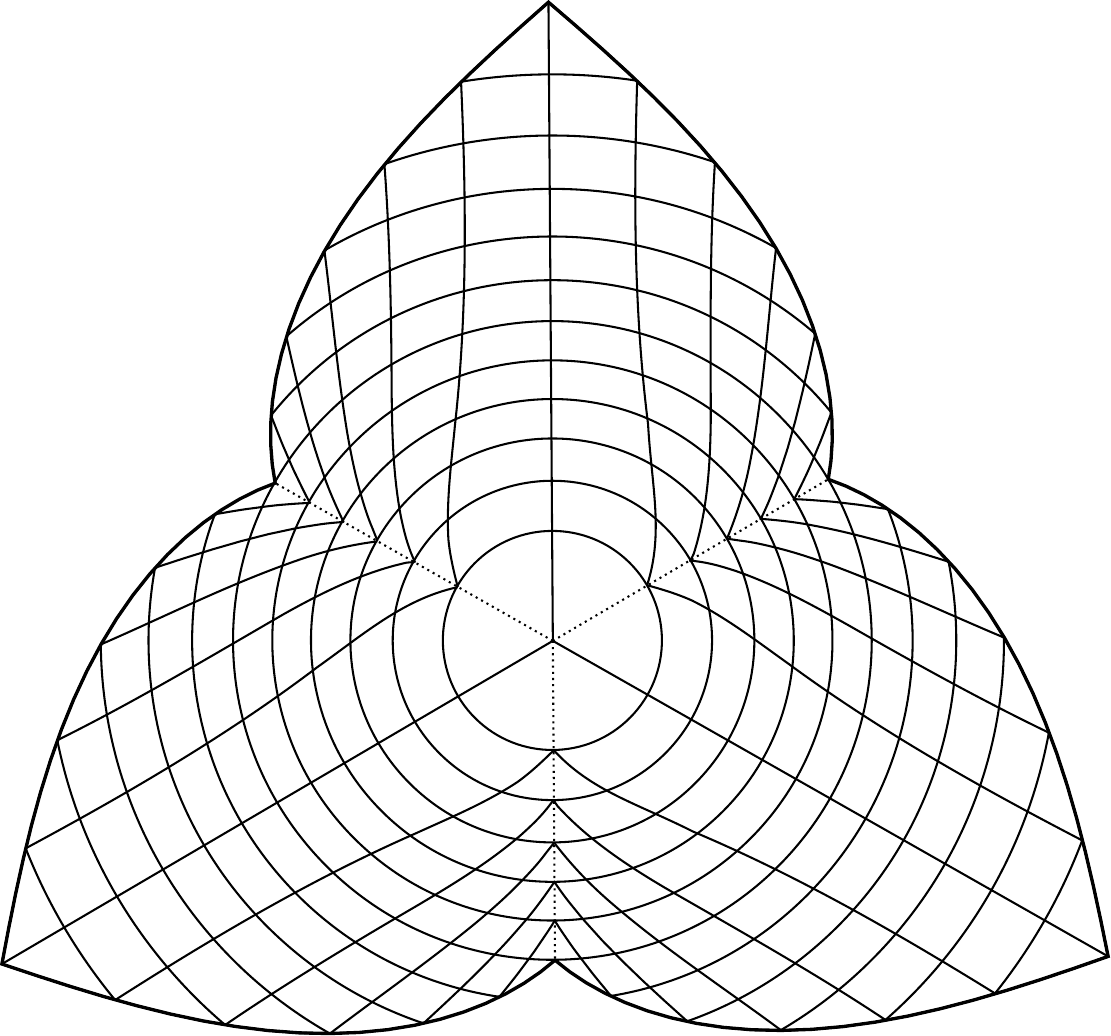}} 
\caption{A visualization of $\mu(x)$ for $n=3$}
\label{fig:figure1}
\end{figure}

In what follows, we describe a somewhat plausible approach. Note that the votes in Approval correspond to vertices of the unit hypercube $I_n=[0,1]^n$. Instead of assuming that the candidates in Score are arbitrary points in $I_n$, we shall impose a slight limitation by assuming that all candidates lie on the facets of $I_n$ that contain $(1,1,\dots,1)$. In other words, we assume that every candidate has at least one coordinate equal to~1. This clearly generalizes Approval where the same assumption means that there are no candidates with $|x|=0$. The normalization function
$$
\nu(x):= \frac{x}{|x|},
$$
used in Approval, maps the $n$ facets containing $(1,1,\dots,1)$ to the simplex $\Delta^{n-1}$. Let $u=(1/n,\dots,1/n)$. Imposing the condition that, on the first step, equal total scores should be treated equally, we replace $\nu(x)$ by $\mu(x)+u$ where
$$
\mu(x)= \frac{\sqrt{\frac{1}{|x|}-\frac{1}{n}}}{||\nu(x)-u||}(\nu(x)-u).
$$
We set $\mu((1,1,\dots,1)):=(0,0,\dots,0)$. This generalizes Approval as it is readily seen that $$\left|\left|\frac{x}{|x|}-u\right|\right|^2=\frac{1}{k}-\frac{1}{n}$$ if $x\in\{0,1\}^n$ has exactly $k$ coordinates equal \mbox{to 1}. See Figure~\ref{fig:figure1} for an illustration of this new normalization function.\\
\vspace*{-.9em}

\par
A method that sequentially chooses $x_{i_1},x_{i_2},\dots$ by minimizing $||\sum_{j}\mu(x_{i_j})||$ generalizes Phragm\'{e}n-Sainte-Lagu\"{e}, but to carry out the improvement with the difference quotients, analogous to the one in Algorithm~\ref{alg:alg2}, we will also assume that the difference $\delta(x,y)$ has at least one coordinate equal to 1 for any two candidates~$x,y$. We can ensure this by adding at least one \emph{phantom voter} per candidate that gives that candidate a score of 1 and a score of 0 to every other candidate. This limits the applications of the method as it is necessary to assume that $n$ is significantly larger than~$m$, so that this modification does not bias the result of the election too much. It is reasonable to assume that $n\gg m$ in parliamentary elections. In such elections, the phantom voters could be the candidates themselves, i.e. the phantom voters could be added and the candidates disfranchised. Alternatively, the phantom voters could be apportioned (based on polls, signatures, previous results etc.) by a divisor method that guarantees at least one phantom voter per candidate.
\par
Before we write down the formal definition, we remark that $\delta$ is still defined as in~\eqref{eq:delta}:
$$\delta(a,b):=(\max\{0,a_k-b_k\})_k,\quad k=1,\dots,n$$
for any $a=(a_1,\dots,a_n)$ and $b=(b_1,\dots,b_n)$. We will need to introduce a new vector to replace $\omega$. To that end, for a finite set $S$ of points on the facets of $I_n$ away from the origin, let
$$
\psi_S:=\sum_{x\in S} \mu(x)
$$
with the convention $\psi_\emptyset=(0,0,\dots,0)$.
\par
\begin{alg}[A generalization of~\ref{alg:alg1}]
\label{alg:alg3}
Let $\mathcal{L}$ be the empty list and let $\mathcal{C}$ be the set of candidates such that every $x\in\mathcal{C}$ has at least one coordinate equal to~1. As long as there are candidates not on the list, append to it the candidate $x\notin\mathcal{L}$ that minimizes
$$
||\psi_{\mathcal{L}}+\mu(x)||.
$$
The list $\mathcal{L}$ gives the ordering of the candidates. Ties are broken by index.
\end{alg}
\par
\begin{alg}[A generalization of~\ref{alg:alg2}]
\label{alg:alg4}
Let $\mathcal{L}$ be the empty list and let $\mathcal{C}$ be the set of candidates such that $\delta(x,y)$ has at least one coordinate equal to~1 for every $x,y\in\mathcal{C}$. As long as there are candidates not on the list, do the following:
\begin{itemize}
\item[1)]Find the candidate $x\notin\mathcal{L}$ for which $||\psi_{\mathcal{L}}+\mu(x)||$ is minimal. 

\item[2)]If the set
$$S_x=\left\{z\in\mathcal{C}\backslash (\mathcal{L}\cup\{x\})\stl ||\psi_{\mathcal{L}}+\mu\comp\delta(z,x)|| < ||\psi_{\mathcal{L}}+\mu\comp\delta(x,z)|| \right\}$$
is empty, then append $x$ to $\mathcal{L}$, otherwise append $y\in S_x$ that maximizes
\begin{equation}
\label{eq:expr2}
\begin{split}
&\left(|| \psi_{\mathcal{L}} + \mu\comp\delta(y,x) + (k+1)u ||^2-|| \psi_{\mathcal{L}} + {k}u ||^2\right)^{-1}+\\
-&
\left(|| \psi_{\mathcal{L}} + \mu\comp\delta(x,y) + (k+1)u ||^2-|| \psi_{\mathcal{L}} + {k}u ||^2\right)^{-1},
\end{split}
\end{equation}
where $k=\#\mathcal{L}$ is the number of candidates elected thus far.
\end{itemize}
The list $\mathcal{L}$ gives the ordering of the candidates. A tie in step 1) is broken by comparing the norms in step 2) that define $S_x$ and if they are equal, the tie is broken by index.
\end{alg}
\begin{rem}
Expression~\eqref{eq:expr2} is chosen so that it generalizes~\eqref{eq:expr1}, which is understood as the difference of the sums of the reweighted votes. This is by no means a canonical choice; many different variants of the method can be defined by choosing a different candidate as the best improvement in the set $S_x$.
\end{rem}
\par
We make several observations about this approach. As we have seen, when restricted to Approval, $\mu(x)={x}/{|x|}-u$ is the difference between the ideal and the representations that the voters receive if $x$ is elected. The sum
$$
\psi_{\mathcal{L}}=\sum_{x\in\mathcal{L}}\mu(x)
$$
of these differences can be interpreted as the overall bias of the elected set $\mathcal{L}$. In Score,~$\mu(x)$ no longer corresponds to an intuitive definition of over-representation, but $\psi_{\mathcal{L}}$ can still be interpreted as a ``bias-vector'' associated to $\mathcal{L}$.\par
\begin{ex}
Let $k$ be a positive integer such that $\dfrac{n}{2}<k<n$ and let
\begin{equation*}
\begin{split}
a&=(\underbrace{1,\dots,1}_\textrm{$k$ times},0,\dots,0),\\[.5em]
b&=\left(\frac{2k-n}{k},\dots,\frac{2k-n}{k},1,\dots,1\right).\\
&\raisebox{1em}{{\hspace{2.4em}$\underbrace{\hspace{7.8em}}_\textrm{$k$ times}    $}}
\end{split}
\end{equation*}
Then $\mu(a)+\mu(b)={0}$. Consider the following classical special case. Let $n=4$ and suppose that there are two seats to apportion and that a candidate $(1,1,1,0)$ is given the first seat. The classical Sainte-Lagu\"{e} method considers that there is a tie for the second seat between a clone $(1,1,1,0)$ and $(0,0,0,1)$. However, both of these choices are suboptimal in the sense that representations
$$
\left(\frac{2}{3},\frac{2}{3},\frac{2}{3},0\right)\;\text{ , }\;\left(\frac{1}{3},\frac{1}{3},\frac{1}{3},1\right)
$$
deviate from the ideal values (by an equal amount but in different directions). However, measuring bias by $\mu$, the ideal candidate for the second seat would be
$$
\left(\frac{2}{3},\frac{2}{3},\frac{2}{3},1\right)
$$
because the two elected candidates would have an equal number of votes ``in opposite directions''.
\end{ex}
\begin{ex}
Let $k$ be a positive integer such that $k\leq\dfrac{n}{2}$ and let
\begin{equation*}
\begin{split}
a&=(1,1,1.\;.\;.\;.\;.\;.\;.\;.\;.\;.\;.\;.\;.\;.\;.\;,1,1),\\[.8em]
b&=(\underbrace{1,1,\;.\;.\;.\;.\;.\;.\;.\;.\;.\;,1}_{2k\text{ times}},0,0,\dots,0),\\
c&=(\underbrace{1,1,\dots,1}_{k\text{ times}},t,t,\;.\;.\;.\;.\;.\;.\;.\;.\;.\;.\;,t),\\
d&=(\underbrace{t,t,\dots,t}_{k\text{ times}},\underbrace{1,1,\dots,1}_{k\text{ times}},t,t,\dots,t),
\end{split}
\end{equation*}
where
$$
t=\frac{3(n-k)}{4n-3k}.
$$
Then $\mu(a)+\mu(b)=\mu(c)+\mu(d)$. In particular, under $\mu$, the two sets
$$
\{(1,1,1),(1,1,0)\}\;,\;\left\{\left(1,\frac{1}{2},\frac{1}{2}\right),\left(\frac{1}{2},1,\frac{1}{2}\right)\right\}
$$
are considered equally biased against the third voter.
\end{ex}
One hopes that examples like these will provide useful insight into the nature of the methods given by Algorithms~\ref{alg:alg3} and~\ref{alg:alg4}.\par

\subsection{Generalizing the new method}
There is more than one way to generalize Algorithms~\ref{alg:new} and~\ref{alg:new2} to Score and we will present two simple generalizations. We change the definition of voter representation in both cases; if $\mathcal{L}$ is the set of candidates elected thus far, we set
\begin{equation}
\label{eq:neww}
{\omega_\mathcal{L}}:=\sum_{x\in\mathcal{L}}\dfrac{x^2}{|x|}.
\end{equation}
\begin{alg}
\label{alg:new_score}
Apply Algorithm~\ref{alg:new} using definitions~\eqref{eq:neww} and~\eqref{eq:newrew}.
\end{alg}
\begin{alg}
\label{alg:new_score2}
Apply Algorithm~\ref{alg:new2} using definitions~\eqref{eq:neww}, \eqref{eq:newrew} and~\eqref{eq:newrewdelta1}.
\end{alg}
The second approach generalizes the reweighting function in a way that is consistent with~\eqref{eq:eq2}. Specifically, for any $y,z,w\in \RR_{\geq 0}^n$, the $k$-th coordinate of~$r(y,w)$ is now defined as
\begin{equation}
\label{eq:newrew2}
r(y,w)_k=
\begin{cases}
			\dfrac{y_k^3}{2w_k{|y|}+y_k^2} & \text{if } y_k \neq 0, \\
      0 & \text{if } y_k = 0.
\end{cases}
\end{equation}
Similarly, we redefine
\begin{equation}
\label{eq:newrewdelta2}
r_\delta(y,z,w)_k=
\begin{cases}
			\dfrac{(y_k-z_k) y_k^2}{2w_k{|y|}+y_k^2} & \text{if } y_k>z_k, \\
      0 & \text{if } y_k \leq z_k.
\end{cases}
\end{equation}
\begin{alg}
\label{alg:new_score3}
Apply Algorithm~\ref{alg:new} using definitions~\eqref{eq:neww} and~\eqref{eq:newrew2}.
\end{alg}
\begin{alg}
\label{alg:new_score4}
Apply Algorithm~\ref{alg:new2} using definitions~\eqref{eq:neww}, \eqref{eq:newrew2} and~\eqref{eq:newrewdelta2}.
\end{alg}

\subsection{Examples}
Finally, to point out the differences between the methods, we provide two examples with scores in $\{0,1/2,1\}$, with phantom voters added. Algorithms~\ref{alg:alg1},~\ref{alg:alg2},~\ref{alg:new} and~\ref{alg:new2} are applied after converting the scores to approval votes via~\eqref{eq:convert}.\par

\begin{ex}For
{\makeatletter
\renewcommand*\env@matrix[1][*\c@MaxMatrixCols c]{
  \hskip -\arraycolsep
  \let\@ifnextchar\new@ifnextchar
  \array{#1}}
\makeatother
\begin{equation*}
\begin{bmatrix}
A\\
B\\
C\\
D\\
E
\end{bmatrix}=\frac{1}{2}
\begin{bmatrix}[ccccc|cccccccccccccccc]
2 & 0 & 0 & 0 & 0 & 0 & 1 & 0 & 1 & 0 & 2 & 2 & 1 & 1 & 1 & 2 & 0 & 2 & 1 & 2 & 0\\
0 & 2 & 0 & 0 & 0 & 2 & 1 & 1 & 2 & 1 & 2 & 1 & 0 & 0 & 1 & 2 & 2 & 2 & 1 & 2 & 0\\
0 & 0 & 2 & 0 & 0 & 2 & 1 & 1 & 1 & 0 & 1 & 1 & 0 & 2 & 0 & 0 & 0 & 0 & 1 & 2 & 1\\
0 & 0 & 0 & 2 & 0 & 1 & 0 & 1 & 0 & 1 & 2 & 2 & 1 & 2 & 0 & 0 & 0 & 2 & 0 & 2 & 0\\
0 & 0 & 0 & 0 & 2 & 0 & 2 & 0 & 1 & 1 & 1 & 1 & 1 & 0 & 1 & 0 & 0 & 2 & 1 & 0 & 2
\end{bmatrix}
\end{equation*}}\par\noindent
we get the following outcomes:

\parbox{.45\textwidth}{
\begin{itemize}
\item Algorithm~\ref{alg:alg1}: $(B,D,E,C,A)$
\item Algorithm~\ref{alg:alg2}: $(B,A,D,E,C)$
\item Algorithm~\ref{alg:new}: $(B,D,E,A,C)$
\item Algorithm~\ref{alg:new2}: $(B,A,E,D,C)$
\item Algorithm~\ref{alg:alg3}: $(B,E,C,A,D)$

\end{itemize}}
\:
\parbox{.45\textwidth}{
\begin{itemize}
\item Algorithm~\ref{alg:alg4}: $(B,A,E,C,D)$
\item Algorithm~\ref{alg:new_score}: $(B,E,D,A,C)$
\item Algorithm~\ref{alg:new_score2}: $(B,A,E,D,C)$
\item Algorithm~\ref{alg:new_score3}: $(B,D,E,A,C)$
\item Algorithm~\ref{alg:new_score4}: $(B,A,C,E,D)$
\end{itemize}}
\end{ex}

\begin{ex}
For
{\makeatletter
\renewcommand*\env@matrix[1][*\c@MaxMatrixCols c]{
  \hskip -\arraycolsep
  \let\@ifnextchar\new@ifnextchar
  \array{#1}}
\makeatother
\begin{equation*}
\begin{bmatrix}
A\\
B\\
C\\
D\\
E
\end{bmatrix}=\frac{1}{2}
\begin{bmatrix}[ccccc|cccccccccccccccc]
2 & 0 & 0 & 0 & 0 & 1 & 0 & 0 & 1 & 0 & 2 & 2 & 2 & 2 & 1 & 1 & 1 & 1 & 2 & 1 & 1\\
0 & 2 & 0 & 0 & 0 & 2 & 2 & 2 & 2 & 1 & 1 & 2 & 1 & 0 & 1 & 0 & 0 & 2 & 1 & 0 & 2\\
0 & 0 & 2 & 0 & 0 & 2 & 1 & 2 & 0 & 2 & 2 & 2 & 0 & 0 & 0 & 2 & 0 & 0 & 1 & 1 & 2\\
0 & 0 & 0 & 2 & 0 & 1 & 2 & 2 & 2 & 0 & 0 & 2 & 0 & 2 & 1 & 1 & 1 & 1 & 2 & 2 & 1\\
0 & 0 & 0 & 0 & 2 & 1 & 2 & 0 & 0 & 2 & 0 & 0 & 2 & 1 & 0 & 1 & 2 & 1 & 0 & 0 & 1
\end{bmatrix}
\end{equation*}}
we get the following outcomes:\par

\parbox{.45\textwidth}{
\begin{itemize}
\item Algorithm~\ref{alg:alg1}: $(D,C,E,B,A)$
\item Algorithm~\ref{alg:alg2}: $(D,B,A,C,E)$
\item Algorithm~\ref{alg:new}: $(D,C,A,B,E)$
\item Algorithm~\ref{alg:new2}: $(D,C,B,A,E)$
\item Algorithm~\ref{alg:alg3}:  $(D,E,A,C,B)$
\end{itemize}}
\:
\parbox{.45\textwidth}{
\begin{itemize}
\item Algorithm~\ref{alg:alg4}:  $(D,A,C,E,B)$
\item Algorithm~\ref{alg:new_score}: $(D,A,C,B,E)$
\item Algorithm~\ref{alg:new_score2}: $(D,A,B,C,E)$
\item Algorithm~\ref{alg:new_score3}: $(D,B,C,E,A)$
\item Algorithm~\ref{alg:new_score4}: $(D,B,C,A,E)$
\end{itemize}}
\end{ex}

{\let\clearpage\relax
}

\bigskip

\noindent
\texttt{djukanovic@gmail.com}
\noindent
Leiden Mathematical Institute, Postbus 9512, 2300 RA Leiden, The Netherlands

\end{document}